\documentclass[a4paper,fleqn]{cas-dc}

\usepackage[font=scriptsize, justification=centering]{caption}
\usepackage{subcaption}
\usepackage[switch]{lineno}
\usepackage{multirow}
\usepackage{enumitem}
\usepackage{listings}
\usepackage{xspace}
\usepackage[flushleft]{threeparttable}
\usepackage[numbers]{natbib}
\usepackage{xcolor}
\usepackage{color,soul}
\usepackage{bibnames}
\usepackage{amssymb}
\usepackage{verbatim}
\usepackage{amsmath}
\usepackage{amsthm}
\usepackage{float}
\usepackage{graphicx}
\usepackage{subcaption}
\usepackage[linesnumbered, ruled,vlined]{algorithm2e}
\usepackage{hyperref}
\usepackage{mdframed}
\usepackage{flushend}
\usepackage{mathrsfs}
\usepackage[T1]{fontenc}

\definecolor{light-gray}{gray}{0.92} 
\newenvironment{gtheorem}%
{\begin{mdframed}[backgroundcolor=light-gray,
skipabove=5pt,
skipbelow=0pt,
nobreak=false
]\begin{mdtheorem}{name}{label}}%
{\end{mdtheorem}\end{mdframed}}

\definecolor{ao}{rgb}{0.0, 0.5, 0.0}
\newcommand{\tool}{\textsc{Rambo}\xspace}

\newcommand{\etal}{\textit{et al.}\space}

\begin{document}


\shorttitle{\tool: Enhancing RAG-based Repo-Level Method Body Completion}

\shortauthors{Bui \textit{et~al.}}

\title [mode = title]{\tool: Enhancing RAG-based Repository-Level Method Body Completion}

\author{Tuan-Dung Bui}
\ead{21020006@vnu.edu.vn}

\affiliation{organization={Faculty of Information Technology, VNU University of Engineering and Technology},
    city={Hanoi},
    country={Vietnam}}

\author{Thieu Luu}
\ead{21020476@vnu.edu.vn}

\author{Thanh-Phat Nguyen}
\ead{22028108@vnu.edu.vn}

\author{Thu-Trang Nguyen}
\ead{trang.nguyen@vnu.edu.vn}

\author{Hieu Dinh Vo}
\ead{hieuvd@vnu.edu.vn}
\cormark[1]

\author{Son Nguyen}
\ead{sonnguyen@vnu.edu.vn}

\cortext[cor1]{Corresponding author}

\begin{abstract}
Code completion is essential in software development, helping developers by predicting code snippets based on context. 
Among completion tasks, Method Body Completion (MBC) is particularly challenging as it involves generating complete method bodies based on their signatures and context. This task becomes significantly harder in large repositories, where method bodies must integrate repository-specific elements such as custom APIs, inter-module dependencies, and project-specific conventions. 
In this paper, we introduce \tool, a novel RAG-based approach for repository-level MBC. Instead of retrieving similar method bodies, \tool identifies essential repository-specific elements, such as classes, methods, and variables/fields, and their relevant usages. By incorporating these elements and their relevant usages into the code generation process, \tool ensures more accurate and contextually relevant method bodies.
Our experimental results with leading code LLMs across 40 Java projects show that \tool significantly outperformed the state-of-the-art repository-level MBC approaches, with the improvements of up to 46\% in BLEU, 57\% in CodeBLEU, 36\% in Compilation Rate, and up to 3X in Exact Match. Notably, \tool surpassed RepoCoder Oracle method by up to 12\% in Exact Match, setting a new benchmark for repository-level MBC.

\end{abstract}


\maketitle

\section{Introduction}

Code completion is a crucial task in software development that assists programmers by suggesting the next units of code based on the current context~\cite{naturalness,bigcode,ase19,arist,tu2014localness}. 
Recently, \textit{Large Language Models for code (Code LLMs)}~\cite{codellama,deepseek-coder,code-gen-app,code-llm-survey} such as Codedex, Code Llama, and Deepseek Coder have emerged as promising tools for code completion, offering the potential to automate repetitive tasks and increase developer productivity.
Several code LLMs have been deployed as auto-completion plugins, such as Github Copilot and CodeGeeX2~\cite{codegeex} in modern IDEs, and successfully streamline real-world software development activities to a certain degree~\cite{reacc,tang2023domain,llm-code-quality,bugs-llm-gen-code}.

One of the most crucial tasks within code completion is \textit{Method Body Completion (MBC)} which is the process of generating the \textit{full} implementation of a method based on its signature and surrounding context.
MBC is particularly important in reducing the manual effort of developers, especially in large-scale projects where method implementations are complex and time-consuming.
Unlike smaller completion tasks such as line completion or API invocation, which involve completing a single line or predicting an API call, MBC requires generating entire method bodies that can span multiple lines of code and may include several API invocations. These method bodies are often unique and require a deeper understanding of the method's intended functionality.
Moreover, \textit{repo(sitory)-level MBC} adds another layer of complexity as it involves requires not only generating method bodies but also integrating repository-specific elements such as user-defined APIs, inter-module dependencies, and project-specific code conventions~\cite{repocoder,tang2023domain,repoformer,reacc}.
Despite its difficulty, repo-level MBC remains an essential task in software development, playing a crucial role in improving developer productivity and maintaining code quality in large projects.

To address the challenges of repo-level code completion, recent approaches have applied the Retrieval-Augmented-Generation (RAG) strategy.
These methods provide repository-specific knowledge to enhance the performance of Code LLMs in completing repo-level code~\cite{repocoder,repoformer,repohyper,rlcoder}. 
RepoCoder~\cite{repocoder} employs a similarity-based retriever in combination with a Code LLM to form an iterative retrieval-generation pipeline. 
This technique retrieves similar code snippets from the repository based on previous code or generated lines, using these examples as additional context for code generation.
Additionally, RepoHyper~\cite{repohyper} and RLCoder~\cite{rlcoder} improve the capability of searching for similar code within RepoCoder's pipeline.
However, useful similar code might not always exist or be found within the repository for code generation. Indeed, Wu~\etal~\cite{repoformer} report that up to 80\% of the standard retrievals performed using RepoCoder's pipeline do \textit{not enhance} the performance of common code LLMs and may \textit{degrade} performance by introducing irrelevant information.
This issue could be more severe for the task of MBC. Method bodies are typically much more unique than smaller completion units such as lines of code (in the Line Completion task) or API invocations (in the API Invocation Completion task). 
Consequently, when provided with dissimilar code, code LLMs could be inaccurately guided by irrelevant context, leading to the generation of low-quality method bodies.

In this paper, we propose \tool, a novel RAG-based approach for \textit{Method Body Completion} (\textit{MBC}). 
Our approach is founded on the idea that constructing a method body correctly in a specific repository requires \textit{\textbf{accurately}} using the \textbf{\textit{essential}} repo-specific code elements.
\textit{Instead of retrieving merely similar method bodies as in existing approaches}, \tool focuses on identifying and retrieving essential repo-specific elements such as declared classes, methods, fields/variables, and their usages relevant to the infilling method. 
These elements serve as the ``\textit{materials},'' and their usages act as the ``\textit{recipes}'' for accurately incorporating these materials into the generated method body.
By providing these essential elements and their relevant usages to a code LLM, \tool enables the model to generate method bodies that are contextually accurate and aligned with the repository's specific requirements.

To identify the essential code elements needed to complete the given infilling method $m$ in a repository, a naive solution might scan the entire codebase for all accessible elements, which would introduce excessive noise. Another approach could focus on methods with similar signatures or contexts; however, these often provide irrelevant elements that do not serve $m$'s functional purpose, leading to redundancy and missing critical elements.
To address this, \tool generates a \textit{sketch} body which provides a rough outline of $m$. This sketch is analyzed to extract the \textit{potentially} essential code elements. These elements are then used to search for similar classes, methods, and fields already declared in the repository which are considered as the essential code elements for constructing $m$'s body.
In \tool, to improve the relevancy of the generated sketch, it is crucial that the sketch code has similar functionality to $m$. In other words, the sketch should use the parameter set to produce a type-correct output with a code structure that generally reflects $m$'s purpose. To achieve this, we generate the sketch by applying RAG with a context that includes methods having a similar signature to $m$ in the current repository.

After identifying the essential code elements, \tool searches for all the methods in the repository that use these elements. These usages are then ranked based on their relevancy to the infilling method $m$. In \tool, the relevancy of an element's usage to $m$ is estimated by the similarity of the usage and $m$. Finally, the most relevant usages of the identified essential code elements are fed into a code LLM to generate $m$'s body.

We conducted several experiments to evaluate the performance of \tool in repo-level method body completion using state-of-the-art code LLMs, including DeepSeek Coder~\cite{deepseek-coder}, Code Llama~\cite{codellama}, and GPT-3.5 Turbo~\cite{gpt}. These experiments were conducted across 40 Java projects with a total of 674 repo-level MBC problems.
Our results demonstrate that \tool significantly surpasses RepoCoder across all evaluated code LLMs and metrics. Specifically, \tool achieves impressive improvements over the other MBC approaches in various performance metrics, by up to 46\% in \textit{BLEU}, up to 57\% in \textit{CodeBLEU}, up to 36\% in \textit{Compilation Rate}, and up to 3X in \textit{Exact Match} rate.
\textbf{Remarkably}, \tool delivers up to 12\% enhancement in  \textit{Exact Match} rate compared to the Oracle RAG-based method~\cite{repocoder}, which estimates the upper-bound of performance for RepoCoder and other methods employing the same pipeline. This substantial improvement shows \tool's superior capability in generating high-quality method bodies and sets a new benchmark in the task of repository-level method body completion.

The contributions of this paper are listed as follows:

\textbf{1. Introduction of \tool}: We present \tool, a novel RAG-based approach specifically designed for repository-level method body completion (MBC), which leverages repo-specific knowledge to enhance code LLM performance.

\textbf{2. Advanced Retrieval Strategy for RAG-based MBC}: We introduce a novel retrieval mechanism that identifies and utilizes essential repo-specific code elements, such as classes, methods, and fields, and their relevant usages to construct more accurate and contextually relevant contexts for RAG-based method body completion.

\textbf{3. Comprehensive Evaluation and Benchmarking}: We conduct extensive experiments with advanced code LLMs, demonstrating that \tool significantly outperforms the state-of-the-art RAG-based MBC techniques, and establish \tool as a new benchmark for MBC.


\section{Problem Statement}

Let $R$ denote a repository containing multiple source files, where each file $f$ includes one or more methods. Each method $m$ is defined by its signature $m_s$ and surrounding context $C_f$, consisting of left context $C_f^{\text{left}}$ (preceding the method) and right context $C_f^{\text{right}}$ (following the method).
The objective of \textit{repository-level Method Body Completion (MBC)} is to generate the method body $\widehat{m_b}$ for a given method $m$, based on its signature $m_s$, context $C_f$, and repository-specific information $R$, formally:
\[
\widehat{m_b} = \arg \max_{B} P(B \mid m_s, C_{f}, R),
\]
where $B$ is a candidate method body, and $P(B \mid m_s, C_f, R)$ is the probability of $B$ being the correct completion, given the method signature, surrounding context, and repository. The key components include:
\begin{itemize}
    \item \textbf{Method Signature} ($m_s$) specifies the method's name, parameters, and return type, guiding the overall structure of the body.
    
    \item \textbf{Surrounding Context} ($C_f$) provides the surrounding code offering insight into the intended behavior and logic of the given method.
    
    \item \textbf{Repository} ($R$) contains project-specific information such as custom APIs, classes, and dependencies.
\end{itemize}

The challenge is to generate a method body $\widehat{m_b}$ that is both compilable and functionally accurate, requiring correctly incorporating of repository-specific elements. This makes repository-level MBC more complex than standard method body completion, which typically focuses on isolated methods without broader dependencies.


\section{Repository-Level Method Body Completion with Retrieval-Augmented Generation}

\begin{figure*}
 \centering
 \includegraphics[width=2.0\columnwidth]{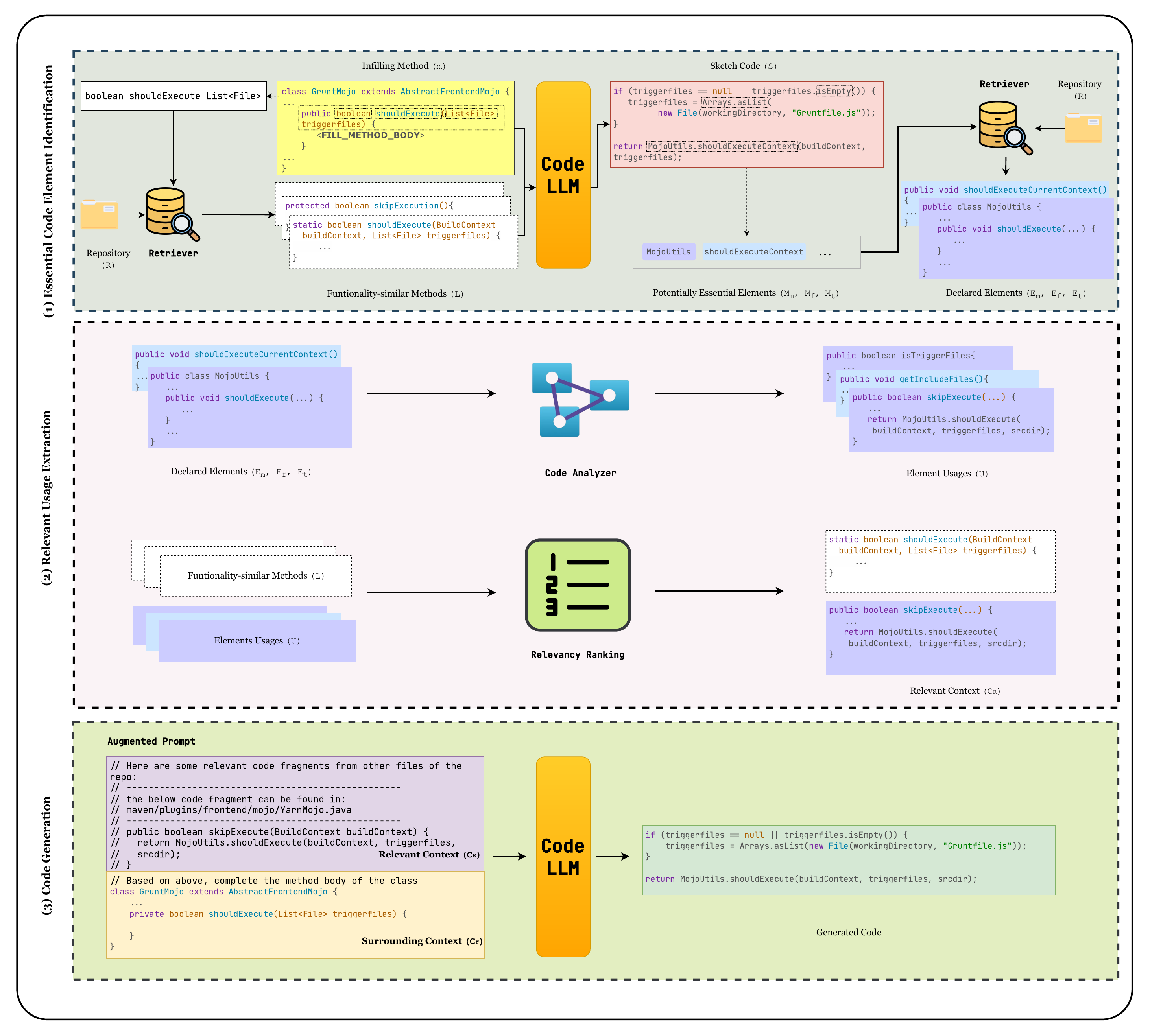}
 \caption{{\tool: A Repository-Level Method Body Completion Approach}}
 \label{fig:approach}
\end{figure*}

Fig.~\ref{fig:approach} shows the overview of our novel RAG-based approach for repo-level MBC, \tool. 
In general, for a given infilling method $m$ in file $f$ of a specific repository $R$, \tool first retrieves for the relevant repo-specific context $C_R$, formally $C_R = \textit{retrieve}(m, R)$. Then, \tool leverage a code LLM, $\mathscr{M}$, to generates the method body $\widehat{m_b}$ utilizing the retrieved relevant context and the surrounding context $C_f$, $\widehat{m_b} = \mathscr{M}(m, C_f, C_R)$.
In \tool, procedure \textit{retrieve} consists of two main steps, \textit{Essential Code Element Identification (EEI)} and \textit{Relevant Usage Extraction (RUE)}.
For the given infilling method $m$ in repo $R$, \textit{EEI} identifies the repo-specific essential code elements (e.g., methods, attributes, or types) as ``materials'' to form $m$'s body (Section~\ref{sec:EEI}). 
After that, \textit{RUE} extracts these elements' usages (i.e., the methods using those elements) as ``recipes'' of these materials to form the method body and then ranks these usages based on the relevancy to $m$ (Section~\ref{sec:RUE}).
Then, the identified essential elements and their relevant usages are utilized to construct the context $C_R$ before feeding $C_R$ into $\mathscr{M}$ to generate method body $\widehat{m_b}$.

\subsection{Repository-specific Context Retrieval}

This section describes the retrieval phase, in \tool to search for the repo-specific context relevant to the given infilling method, $C_R = \textit{retrieve}(m, R)$. Our retrieval approach consists of two phases, \textit{Essential Code Element Identification} and \textit{Relevant Usage Extraction}.

\subsubsection{Essential Code Element Identification (EEI)} \label{sec:EEI}

In the \textit{Essential Code Element Identification} (\textit{EEI}) phase, the goal is to identify the code elements within repo $R$ that are crucial for constructing the given method, $m$. 
The challenge here is to ensure that these repo-specific elements are contextually relevant, meaning they should align with the specific needs of the method being completed while adhering to the broader repository's structure and dependencies. 

A \textit{naive solution} might involve scanning the entire codebase to identify all accessible code elements from the completion point. 
This could lead to an overwhelming amount of information, much of which may be irrelevant, thereby providing much noise to the code generation process. 
Another approach is to search for methods similar to $m$'s signature $m_s$ or having a similar surrounding context $C_f$. The elements used in these similar methods could then be extracted as essential candidates for constructing $m$'s body. 
However, the methods that are similar in the signature or surrounding content may not necessarily provide the right elements for $m$, especially if they serve different functional purposes within the repository. Consequently, the resulting set of elements could still contain many redundant or irrelevant items while potentially missing key elements that are critical for the correct construction of $m$'s body.

\textit{In the EEI process of }\tool, our idea is to leverage the capabilities of code LLMs in understanding and generating code to narrow down the set of elements essential for method body construction. 
Specifically, we employ $\mathscr{M}$ to generate a \textit{sketch} body that provides a rough outline of the infilling $m$. This sketch serves as a preliminary framework for understanding the likely structure and components of $m$.
Based on the elements used in this sketch (such as method calls, field accesses, and type declarations), \tool then searches the repository $R$ for similar elements that have already been declared. In \tool, these similar elements are considered essential for forming $m$'s body because they align with $R$'s structure and the functional needs of $m$.

To enhance the relevance of the sketch, \tool first searches for the set of methods, $L$, within the repository $R$ whose interface (input-output) and name are most similar to those of $m$. The set of methods $L$ serves as the reference points for understanding the expected structure and functionality of $m$. 
Then, \tool employs a RAG-based method to generate a sketch code $\mathcal{S}$ of the method body using the retrieved methods $L$ and the surrounding context $C_f$. Although the sketch, generated as $\mathcal{S} = \mathscr{M}(m, L, C_f)$, may not fully accurate, $\mathcal{S}$ includes potentially essential elements such as called methods, used fields, and types. These elements are then analyzed and extracted:
$$
\{\mathcal{M}_m, \mathcal{M}_f, \mathcal{M}_t\} = \text{analyze}(\mathcal{S})
$$
where $\mathcal{M}_m$, $\mathcal{M}_f$, and $\mathcal{M}_t$ are the sets of called methods, accessed fields, and used types in the sketch $\mathcal{S}$, respectively.
Note that analyzing the generated sketch code $\mathcal{S}$ by existing code analyzers (e.g., JDT or Joern~\cite{joern}) is feasible because the code generated by pre-trained code models, especially code LLMs, is quite syntactically correct~\cite{pretrain-model-code-syntax,llm-code-quality}. However, the employed code LLM may hallucinate code, possibly generate the code elements not existing/declared in the repo, or use them inappropriately~\cite{bugs-llm-gen-code,llm-code-quality}. 

To mitigate code LLM hallucination, after analyzing, \tool searches for similar code elements already declared within the repository. This involves identifying the sets of methods, fields, and classes/types whose names are most similar to those found after analyzing $\mathcal{S}$:

\begin{equation}
\label{eq:Em}
E_m = \bigcup_{\substack{
  e \in \mathcal{M}_m
  }} \left\{\arg\max\limits_{m' \in \text{med}(R)} \left(\text{name\_sim}(e, m')\right)\right\}
\end{equation}

\begin{equation}
\label{eq:Ef}
E_f = \bigcup_{\substack{
  e \in \mathcal{M}_f
  }} \left\{\arg\max\limits_{v \in \text{fid}(R)} \left(\text{name\_sim}(e, v)\right)\right\}
\end{equation}

\begin{equation}
\label{eq:Ec}
E_t = \bigcup_{\substack{
  e \in \mathcal{M}_t
  }} \left\{\arg\max\limits_{c \in \text{cls}(R)} \left(\text{name\_sim}(e, c)\right)\right\}
\end{equation}
where $E_m$, $E_f$, and $E_t$ are the sets of accessible methods, fields, and classes whose names are most similar to the methods, fields, and types in $\mathcal{M}_m$, $\mathcal{M}_f$, and $\mathcal{M}_t$, respectively. Also, the functions $\text{med}(R)$ (in Formula \ref{eq:Em}), $\text{fid}(R)$ (in Formula \ref{eq:Ef}), and $\text{cls}(R)$ (in Formula \ref{eq:Ec}) return the sets of all the accessible methods, fields, and classes existing in $R$.


\subsubsection{Relevant Usage Extraction (RUE)} \label{sec:RUE}

In the \textit{Relevant Usage Extraction} (\textit{RUE}) phase, \tool focuses on extracting the relevant usages of these identified elements. The intuition here is that understanding how these elements are used within the repository can provide code LLMs valuable insights into how they should be utilized within $m$. By analyzing the repository, \tool identifies methods that make use of these essential elements:
$$
U = \bigcup_{\substack{
  e \in E_m \cup E_f \cup E_t
  }} \left\{m' \in \text{med}(R) | m' \text{ uses } e\right\}
$$ 
Once these usages are identified, they are ranked based on their relevancy to the functionality of the method $m$. The relevancy of a usage $u$ is calculated by function $\text{rel}$:
$$
\text{rel}(u) \approx \text{sim}(u, \mathcal{S})
$$
where $\text{sim}(u, \mathcal{S})$ calculates the similarity between a usage $u \in U$ and the sketch $\mathcal{S}$ generated during \textit{EEI}. 
The reason for this estimation of $\text{rel}(u)$ is that the body of $u$ expresses how an essential element $e$ is utilized and $\mathcal{S}$ roughly expresses how potentially essential elements should be used.
Thus, if the usage of $e$ is also present in the sketch code, it indicates that the usage of $e$ in $u$ is likely relevant to $m$.
This ranking ensures that the most relevant usages are prioritized, providing the code LLM with high-quality context that is both specific to the repository and relevant to the task at hand.
Additionally, \tool brings back the set of methods $L$ whose signature most closely matches that of $m$, serving as a structural and functional reference point for method $m$.
The methods in $L$ and the top relevant usages are concatenated to form the repo-specific relevant context $C_R$. Finally, this context and the surrounding context $C_f$ of $m$ are fed into $\mathscr{M}$ to generate the final method body $\widehat{m_b} = \mathscr{M}(m, C_f, C_R)$.

\textbf{Discussion.} In the \textit{EEI} phase of the current \tool's version, a \textit{single sketch} is generated for each method to identify essential material. In fact, \textit{multiple sketches} could be generated for each method to increase the likelihood that the most relevant and diverse code elements are identified for method body construction. By generating multiple sketches, \tool can explore various plausible ways in which the method $m$ could be completed within the repository $R$. Each sketch provides a different perspective on how the essential elements, such as method calls, field usages, and type initialization, can be utilized to form the method body.
Additionally, this multi-sketch approach enhances the robustness of the retrieval process by allowing \tool to capture a broader range of potential code structures and usage patterns that may be relevant to $m$. This mitigates the risk of missing out on critical context that could be overlooked if only a single sketch were generated.

However, generating and processing multiple sketches could introduce significant computational overhead, particularly in large repositories or under time constraints. Moreover, it increases the potential for redundancy and noise due to the retrieval process with irrelevant elements. Given these trade-offs, the current version of \tool adopts a simpler single-sketch approach.


\subsection{Method Body Generation}
In \tool, the code generation phase synthesizes the method body $\widehat{m_b}$ for an infilling method $m$ using the retrieved context $C_R$ and given surrounding context $C_f$. This process leverages a large language model specialized in code generation, denoted as $\mathscr{M}$, to produce repository-specific method bodies that align with the existing codebase.
Particularly, the method body $\widehat{m_b}$ is generated by passing the method $m$ and context $C$ into the code LLM: $\widehat{m_b} = \mathscr{M}(m, C_R, C_f)$. 

To generate $m$'s body, we design a prompt template that seamlessly integrates the identified relevant context $C_R$ with the infilling method, embedding the context as code comments within the file (as illustrated in Fig.~\ref{fig:approach}). The retrieved usages are arranged in ascending order of relevancy, with each code snippet annotated by its original file path. The number of snippets included in the prompt is optimized based on the available prompt length. This approach ensures that the prompt is packed with relevant information, facilitating the generation of a method body that is both functionally accurate and consistent with the repository's coding standards.


\section{Evaluation Methodology}

\label{sec:eval}
To evaluate the repo-level method body completion (MBC) performance of our approach, we seek to answer the following research questions:

\noindent\textbf{RQ1: \textit{Accuracy and Comparison}.} How accurate is {\tool} in completing method bodies with the state-of-the-art code LLMs~\cite{gpt,deepseek-coder,codellama}? And how is it compared to the state-of-the-art repo-level MBC approaches~\cite{repocoder}?

\noindent\textbf{RQ2: \textit{Intrinsic Analysis}.} How do different aspects of our approach impact the MBC performance of \tool, including EEI and RUE phases, code embedding techniques used to compute similarity between code, and code LLM's size?

\noindent\textbf{RQ3: \textit{Sensitivity Analysis}.} How do various input factors, including surrounding context length and repository size, affect {\tool}'s performance?

\noindent\textbf{RQ4: \textit{Time Complexity}.} What is  {\tool}'s running time?

\subsection{Dataset}

In this work, we evaluate the MBC performance of \tool and other approaches on two large MBC benchmarks (Table~\ref{tab:dataset-overview}).
First, we utilize Defects4J~\cite{defects4j} as in CoderUJB~\cite{CoderUJB}, a multi-task benchmark for several SE tasks such as functional code generation, test generation, defect detection, and automated program repair. 
In this work, we adapt the functional code generation task from Defects4J in CoderUJB to the MBC task, resulting in 204 MBC problems in 14 real open-source Java projects. 
Additionally, to evaluate the performance of the repo-level MBC approaches on more up-to-date repositories, we construct another benchmark which includes active high-quality open-source repositories from GitHub. Specifically, we collect 26 Java repositories (other than those in Defects4J) that meet the following criteria: they must have an open-source license, have been actively updated, possess over 2K stars, and be compilable. 
From each non-test class in the collected projects, we randomly select a method to generate an MBC problem and the corresponding reference, which is the method's body. To ensure a reasonable difficulty level of the benchmark, we exclude all constructors, \texttt{getters}, \texttt{setters}, and too-short methods with fewer than 3 LOCs. As a result, we obtain a total of 470 MBC problems.
In this work, for every MBC problem, we omitted the import section in the left context, $C_f^{\text{left}}$, to accurately reflect the coding activities in practice, that the necessary imports are usually unavailable before implementing the infilling method.

\begin{table}[]
\scriptsize
\centering
\caption{Datasets statistics}
\begin{tabular}{lrr}
\toprule
                            & \tool's \textbf{benchmark}    & \textbf{Defects4J} \\
\midrule
\#Repositories              & 26                            & 14    \\
\#Files/repository          & [49 -- 4,061]                 & [11 -- 813]   \\
\#Methods/repository        & [87 -- 43,601]                & [537 -- 25,476]   \\
Last update  & May 2024      & July 2019   \\
\#MBC problems              & 470                           & 204   \\
Body size in LOCs           & [3 -- 105]                    & [5 -- 45] \\
Body size in code tokens    & [25 -- 1,434]                 & [45 -- 395]   \\
Test availability           & NO                            & YES   \\

\bottomrule
\end{tabular}
\label{tab:dataset-overview}
\end{table}


\subsection{Evaluation Setup, Procedure, and Metrics}
\subsubsection{Empirical Procedure} 

\textbf{RQ1. Accuracy and Comparison}. 

\textbf{\textit{Baselines.}} We compared \tool against the state-of-the-art repo-level MBC approaches: 

\begin{enumerate}
    \item \textit{In-File Completion}~\cite{codegen,codet}: The conventional In-File Completion (\textit{In-File} for short) utilizes large code language models for code generation in a zero-shot completion manner, conditioned on the surrounding context, including both left and right ones.

    \item {\textit{RawRAG}: This baseline employs the RAG strategy in its most straightforward form by retrieving relevant context solely based on the left context using a similarity search. The retrieved snippets are then directly fed into the code LLM for generation without further refinement of the retrieved content.}

    \item {\textit{RLCoder}}~\cite{rlcoder}{: RLCoder is a reinforcement learning framework that iteratively learns to retrieve high-quality context. In our work, RLCoder has been adapted to generate method bodies line by line until a complete body is formed.}
     
    \item \textit{RepoCoder}~\cite{repocoder}: RepoCoder employs a similarity-based retriever and a code LLM in an iterative retrieval-generation pipeline to generate complete unfinished code, e.g., unfinished lines, API invocations, or function/method bodies.

    \item \textit{RepoCoder Oracle}~\cite{repocoder}: This method estimates the upper bound of RepoCoder's performance by utilizing the ground truth code to search for relevant context before feeding into code LLMs in the RepoCoder. Note that this method performs a single retrieval process to obtain relevant code snippets. 
    
    \item \tool \textit{Oracle}: Inspired by RepoCoder Oracle~\cite{repocoder}, we estimated the upper bound of \tool's performance by its own Oracle. Within \tool's pipeline, this oracle method directly considers the ground truth code as the sketch code $\mathcal{s}$ in EEI and RUE. In this case, the essential elements are correctly identified, and their usages are ranked based on the similarities with the ground truth.
 
\end{enumerate}
We used the implementation in their original papers for all the baseline approaches. For RepoCoder and RepoCoder Oracle, we used the hyper-parameters recommended in the original work~\cite{repocoder}. Additionally, we applied RepoCoder with a single iteration at which RepoCoder achieves the best overall performance with our studied dataset.
We evaluated the MBC performance of the approaches under study using several representative code LLMs, both open-source and commercial, including DeepSeek Coder 6.7B~\cite{deepseek-coder}, Code Llama 7B~\cite{codellama}, and GPT-3.5 Turbo~\cite{gpt} with about 175B parameters.
For detailed implementation and other hyper-parameters, one can see our website~\cite{website}. Note that all experiments were run on a machine with an Intel Xeon Bronze 3104 (6 cores) @ 1.70GHz, 32GB RAM, and an NVDIA RTX A6000 GPU.

\textbf{RQ2. Intrinsic Analysis}. We conducted several experiments to study the impact of \tool's components on the MBC performance of \tool. We analyzed the impact of the following aspects:

\begin{itemize}

    \item \textit{Retrieval Technique}: We evaluated how the retrieval technique powered by different embedding methods affects the MBC performance of \tool.
    
    \item \textit{EEI and RUE: Contribution Analysis}: We examined how \textit{EEI} and \textit{RUE} contribute to the overall repo-level MBC performance of \tool.
    
    \item \textit{Relevant Context Size}: We analyzed how changes in the size of the relevant context, $C_R$, influence the MBC performance.

    \item \textit{Code LLM's Size}: We studied the effect of the size of the employed code LLM, $\mathscr{M}$, on \tool's performance, balancing computational efficiency and the quality of the generated method bodies.
\end{itemize}

\textbf{RQ3. Sensitivity Analysis}. We investigated how variations in input factors affect the performance of \tool. Specifically, we focused on:

\begin{itemize}
    \item \textit{Surrounding Context Size}: We analyzed how changes in the size of the left and right contexts influence the MBC performance of \tool.

    \item \textit{Repository Size}: We examined the impact of repository sizes on \tool's ability to retrieve relevant context and generate method bodies.
\end{itemize}

\subsubsection{Metrics}
To evaluate the performance of \tool and other approaches, we employed several metrics to assess the quality of the generated bodies across various dimensions.

\begin{itemize}

\item \textbf{BLEU}~\cite{bleu} is a widely used metric in NLP, assessing the similarity between generated and reference text by comparing overlapping \textit{n}-grams. In the context of MBC, it measures surface-level similarity, focusing on the direct overlap of tokens between the generated and reference code.

\item \textbf{CodeBLEU}~\cite{codebleu} extends the BLEU score by incorporating code-specific properties, such as syntax and semantic structure. It measures the alignment between the generated and the reference codes, capturing deeper structural elements. 

\item \textbf{Compilation Rate (CR)} assesses the percentage of generated code bodies that successfully compile without errors. A higher compilation rate reflects the syntactical correctness of the generated code and its adherence to the language's rules, making it a crucial metric for evaluating the practicality of code generation models.

\item \textbf{pass@\textit{k}}~\cite{HumanEval} evaluates the generated code's ability to pass a set of unit tests, measuring the percentage of solutions that pass all the given tests when considering the top-\textit{k} predictions. A higher pass@\textit{k} value indicates stronger model performance in producing functionally correct code. In this work, $k=1$ and \textit{pass@k} is applicable for evaluation using Defects4J.

\item \textbf{Exact Match (EM)} quantifies the rate at which the generated method body exactly matches the actual method body. This metric is akin to string equality checks in many programming languages, where both case and spacing are considered.

\end{itemize}


\section{Experimental Results}
\label{sec:results}

\subsection{Answering RQ1: Accuracy and Comparison}

\begin{table}[]
\centering
\caption{{Repo-level MBC performance comparison on \tool's benchmark}}
\label{tab:comparison_rambo}
\resizebox{\columnwidth}{!}{%
\begin{tabular}{l|l|rrrr}\toprule

\textbf{Code LLM} & &\textit{BLEU} &\textit{CodeBLEU} &\textit{CR} &\textit{EM} \\\midrule

\multirow{5}{*}{CodeLlama (7B)} &\cellcolor[HTML]{ffe599}\tool Oracle &\cellcolor[HTML]{ffe599}41.83 &\cellcolor[HTML]{ffe599}46.22 &\cellcolor[HTML]{ffe599}51.28 &\cellcolor[HTML]{ffe599}14.89 \\
&\cellcolor[HTML]{ffe599}RepoCoder Oracle &\cellcolor[HTML]{ffe599}46.59 &\cellcolor[HTML]{ffe599}49.53 &\cellcolor[HTML]{ffe599}55.96 &\cellcolor[HTML]{ffe599}14.68  \\
    &In-File &25.09 &33.85 &48.09 &3.83 \\
    &RawRAG &41.57	&45.96	&54.89	&12.55 \\
    &RLCoder &39.92	&44.30	&54.47	&10.85 \\
    &RepoCoder &42.98 &47.66 &55.32 &12.77 \\
    &\tool &\textbf{43.16} &\textbf{47.76} &\textbf{55.96} &\textbf{13.62} \\\cmidrule{1-6}
\multirow{5}{*}{DeepSeek Coder (6.7B)} &\cellcolor[HTML]{ffe599}\tool Oracle &\cellcolor[HTML]{ffe599}50.87 &\cellcolor[HTML]{ffe599}53.96 &\cellcolor[HTML]{ffe599}63.40 &\cellcolor[HTML]{ffe599}18.72  \\
&\cellcolor[HTML]{ffe599}RepoCoder Oracle &\cellcolor[HTML]{ffe599}50.04 &\cellcolor[HTML]{ffe599}52.48 &\cellcolor[HTML]{ffe599}58.51 &\cellcolor[HTML]{ffe599}17.02 \\
    &In-File &25.99 &33.84 &45.11 &4.04 \\
    &RawRAG &46.92	&50.97	&60.21	&15.32 \\
    &RLCoder &45.08	&48.58	&55.96	&14.26 \\
    &RepoCoder &46.22 &50.31 &57.23 &15.53 \\
    &\tool &\textbf{50.44} &\textbf{53.20} &\textbf{61.70} &\textbf{18.94} \\\cmidrule{1-6}
\multirow{5}{*}{GPT3.5 - Turbo (175B)} &\cellcolor[HTML]{ffe599}\tool Oracle &\cellcolor[HTML]{ffe599}51.21 &\cellcolor[HTML]{ffe599}53.91 &\cellcolor[HTML]{ffe599}60.85 &\cellcolor[HTML]{ffe599}19.15 \\
&\cellcolor[HTML]{ffe599}RepoCoder Oracle &\cellcolor[HTML]{ffe599}47.82 &\cellcolor[HTML]{ffe599}51.43 &\cellcolor[HTML]{ffe599}62.13 &\cellcolor[HTML]{ffe599}17.02  \\
    &In-File &33.93 &39.25 &53.11 &6.38 \\
    &RawRAG &42.90	&45.27	&55.53	&12.55 \\
    &RLCoder &45.12	&49.48	&58.72	&15.32 \\
    &RepoCoder &44.41 &49.64 &58.09 &15.74  \\
    &\tool &\textbf{49.58} &\textbf{52.71} &\textbf{58.94} &\textbf{19.15} \\
\bottomrule
\end{tabular}
}
\end{table}

\begin{table}[]\centering
\caption{{Repo-level MBC performance comparison on Defect4J dataset}}
\label{tab:comparison_defect4j}
\resizebox{\columnwidth}{!}{%
\begin{tabular}{l|l|rrrrrr}\toprule

\textbf{Code LLM} & &\textit{BLEU} &\textit{CodeBLEU} &\textit{CR} &\textit{pass@k} &\textit{EM} \\\midrule

\multirow{5}{*}{CodeLlama (7B)} &\cellcolor[HTML]{ffe599}\tool Oracle &\cellcolor[HTML]{ffe599}54.19 &\cellcolor[HTML]{ffe599}55.90 &\cellcolor[HTML]{ffe599}64.71 &\cellcolor[HTML]{ffe599}46.08 &\cellcolor[HTML]{ffe599}19.12 \\
&\cellcolor[HTML]{ffe599}RepoCoder Oracle  &\cellcolor[HTML]{ffe599}53.69 &\cellcolor[HTML]{ffe599}54.47 &\cellcolor[HTML]{ffe599}64.71 &\cellcolor[HTML]{ffe599}46.57 &\cellcolor[HTML]{ffe599}19.61 \\
&In-File &44.06 &46.73 &55.88 &33.82 &8.82 \\
&RawRAG & 48.88 & 50.93&60.29&45.10&18.01\\
&RLCoder & 50.37&51.12&60.29&42.65&17.54\\
&RepoCoder  &51.27 &53.46 &60.78 &\textbf{46.57} &20.10 \\
&\tool &\textbf{52.04} &\textbf{54.98} &\textbf{63.73} &44.12 &\textbf{20.59} \\\midrule
\multirow{5}{*}{DeepSeek Coder (6.7B)} &\cellcolor[HTML]{ffe599}\tool Oracle  &\cellcolor[HTML]{ffe599}51.31 &\cellcolor[HTML]{ffe599}53.54 &\cellcolor[HTML]{ffe599}67.16 &\cellcolor[HTML]{ffe599}49.51 &\cellcolor[HTML]{ffe599}21.57 \\
&\cellcolor[HTML]{ffe599}RepoCoder Oracle  &\cellcolor[HTML]{ffe599}50.74 &\cellcolor[HTML]{ffe599}52.96 &\cellcolor[HTML]{ffe599}64.71 &\cellcolor[HTML]{ffe599}45.59 &\cellcolor[HTML]{ffe599}21.57 \\
&In-File &39.18 &43.24 &52.45 &34.80 &11.27 \\
&RawRAG &44.96&48.47&61.27&45.10&18.01\\
&RLCoder&45.40&48.10&57.84&45.59&18.01\\
&RepoCoder &48.39 &51.01 &60.78 &46.08 &20.59 \\
&\tool  &\textbf{51.03} &\textbf{53.05} &\textbf{69.61} &\textbf{46.57} &\textbf{23.04} \\\midrule
\multirow{5}{*}{GPT3.5 - Turbo (175B)} &\cellcolor[HTML]{ffe599}\tool Oracle  &\cellcolor[HTML]{ffe599}67.80 &\cellcolor[HTML]{ffe599}67.96 &\cellcolor[HTML]{ffe599}76.96 &\cellcolor[HTML]{ffe599}61.76 &\cellcolor[HTML]{ffe599}35.29 \\
&\cellcolor[HTML]{ffe599}RepoCoder Oracle &\cellcolor[HTML]{ffe599}62.76 &\cellcolor[HTML]{ffe599}64.85 &\cellcolor[HTML]{ffe599}72.06 &\cellcolor[HTML]{ffe599}57.84 &\cellcolor[HTML]{ffe599}31.37 \\
&In-File &63.16&65.66 &\textbf{76.96} &60.78 &26.96 \\
&RawRAG&66.09&67.36&75.00&61.27&31.28\\
&RLCoder&65.78&66.67&75.98&61.76&\textbf{32.84}\\
&RepoCoder &63.52 &65.39 &74.02 &59.80 &31.86 \\
&\tool &\textbf{66.29} &\textbf{67.82} &76.47 &\textbf{63.73} &\textbf{32.84} \\
\bottomrule
\end{tabular}
}
\end{table}

Table~\ref{tab:comparison_rambo}  and Table~\ref{tab:comparison_defect4j} present the MBC performance of \tool compared to the state-of-the-art approaches using several representative code LLMs across our proposed benchmark and the Defects4J. 
As seen, \tool exhibits significant improvements over In-File, RawRAG, RLCoder~\cite{rlcoder}, RepoCoder~\cite{repocoder}, and even RepoCoder Oracle. Even when compared to its own Oracle (\tool Oracle), \tool maintains a highly competitive performance.
%
%
For example, with GPT-3.5 Turbo on \tool's benchmark, \tool achieves a significant increase of about \textbf{12--46\%} in BLEU and \textbf{6--34\%} in CodeBLEU compared to those of RepoCoder, RLCoder, RawRAG, and In-File. Similarly, on the Defects4J dataset with Code Llama, \tool shows consistent gains, outperforming the other approaches by \textbf{2--18\%} in BLEU and \textbf{3--17\%} in CodeBLEU. These results highlight \tool's ability to consistently generate syntactically accurate and high-quality code across diverse repositories, outperforming RepoCoder and RLCoder by a substantial margin in both BLEU and CodeBLEU metrics.
This shows \tool's ability to generate syntactically and semantically accurate code with fewer errors.

Moreover, \tool excels in generating method bodies that is compilable and exactly match the reference implementations, as shown by its significantly better \textit{CR (Compilation Rate)} and \textit{Exact Match} (\textit{EM}) rates. With DeepSeek Coder on \tool's benchmark, \tool's EM rate improves \textbf{22\%} and \textbf{3X} the corresponding figures of RepoCoder and In-File, respectively. With the same code LLM, \tool delivers substantial \textit{CR} improvements of \textbf{8\%}--\textbf{15\%} compared to RepoCoder in Defects4J. In terms of \textit{pass@k}, \tool also outperforms with 2/3 studied code LLMs, with the improvements of \textbf{1--33\%}, further affirming its effectiveness in generating functionally correct code.

Especially our results indicate that \tool not only achieves significantly better performance compared to RepoCoder but also surpasses RepoCoder Oracle, which could represent the upper bound of RepoCoder's potential performance and those of other similarity-based approaches. 
This shows that \tool is not merely an incremental improvement but a substantial advancement in the task of repository-level method body completion.

\begin{gtheorem}
\textbf{Answer to RQ1}: Our results show \tool's consistent superiority over existing repo-level MBC approaches and its competitive performance compared to \tool Oracle. \tool not only outperforms existing approaches but also sets a new benchmark by exceeding the anticipated upper-bound performance of RepoCoder, showing its robustness, accuracy, and practical utility in real-world code generation tasks. 
\end{gtheorem}

\textbf{Result Analysis}. %
%
%
%
%
Upon analyzing the results, we observed that the quality of method bodies generated by \tool is highly dependent on the accuracy of the initial sketch code. When the sketch code is poorly generated, it fails to narrow down the set of essential code elements effectively. This leads to the misidentification of essential elements (in EEI) and irrelevant extraction of usages (in RUE), ultimately providing the code LLM with inadequate context, resulting in low-quality method bodies. A similar issue arises in RepoCoder when the code generated in the first iteration is suboptimal; subsequent iterations rarely see significant improvements. This problem is particularly pronounced in repositories with highly unique and distinct method implementations.

Fig.~\ref{fig:good-example} shows an MBC problem in repository \texttt{Spring Cloud Gateway}\footnote{\url{https://github.com/spring-cloud/spring-cloud-gateway}} with the infilling method \texttt{delete} simplified in Fig.~\ref{fig:MBC_problem}. 
The \texttt{delete} method in the \texttt{InMemoryRouteDefinition-} \texttt{Repository} class removes a \texttt{RouteDefinition} by its ID. It performs a \textit{Route Existence Check} by verifying if the route exists in the map. 
If found, the method removes it and returns an empty \texttt{Mono<Void>} (\textit{Success Case}). If the route ID is not present, it uses \texttt{Mono.defer} to lazily create and return a \texttt{Mono.error} emitting a \texttt{NotFoundException} (\textit{Error Handling}).
In this case, since the other declared methods in the repository are quite different from \texttt{delete}, the context provided to RepoCoder and its Oracle is insufficient. As shown in Fig~\ref{fig:repocoder-code}, the body generated by RepoCoder and its Oracle completely omits both the \textit{Route Existence Check} and \textit{Error Handling}.

On the other hand, the sketch code generated by \tool (Fig.~\ref{fig:rambo-sketch}) includes the \textit{Route Existence Check} and \textit{Success Case}, but still \textbf{misinterprets} the error handling. Based on the sketch, \tool identifies the essential elements and their usages within the repository. One relevant usage is the \texttt{save} method in the \texttt{RedisRouteDefinitionRepository} class  (Fig.~\ref{fig:rel-usage}).
With this relevant usage as a reference where \texttt{Mono.error} is used as expected (lines 8--9 in \texttt{save}), \tool and the code LLM correctly generate the expected body for this MBC problem, as shown in Fig.~\ref{fig:rambo-code}.

\begin{figure}
    \begin{subfigure}{0.5\textwidth}
\begin{lstlisting}[language=Java]
class InMemoryRouteDefinitionRepository {
   Map<String, RouteDefinition> routes = ...;
   Mono<Void> save(Mono<RouteDefinition> route) {
       return route.flatMap(r -> {
           ...
           routes.put(r.getId(), r);
           return Mono.empty();
       });
   }

   Mono<Void> delete(Mono<String> routeId) {
        /* <FILL_METHOD_BODY> */
   }
   ...
}
\end{lstlisting}
        \caption{Simplified MBC problem to fill in the body of \texttt{delete}}
        \label{fig:MBC_problem}
    \end{subfigure}\hfil 
    
    \begin{subfigure}{0.5\textwidth}
\begin{lstlisting}[language=Java]
        return routeId.flatMap(id -> { 
            routes.remove(id); 
            return Mono.empty(); 
        });
\end{lstlisting}
        \subcaption{Method body generated by RepoCoder and RepoCoder Oracle with DeepSeek Coder are identical}
        \label{fig:repocoder-code}
    \end{subfigure}\hfil

    \begin{subfigure}{0.5\textwidth}
\begin{lstlisting}[language=Java]
        return routeId.flatMap(id -> { 
            if (routes.containsKey(id)) { 
                routes.remove(id); 
                return Mono.empty(); 
            } 
            return Mono.error(new NotFoundException("...")); 
        });
\end{lstlisting}
        \subcaption{Sketch code generated to identify essential code elements and their relevant usages by \tool}
        \label{fig:rambo-sketch}
    \end{subfigure}\hfil

    \begin{subfigure}{0.5\textwidth}
\begin{lstlisting}[language=Java]
class RedisRouteDefinitionRepository{
    ...
    Mono<Void> save(Mono<RouteDefinition> route) {
         return route.flatMap(routeDefinition -> routeDefinition... {
            if (success) {
                return Mono.empty();
            }
            return Mono.defer(() -> Mono.error(new 
             RuntimeException(String.format("..."))));
        }));
    }
    ...
}
\end{lstlisting}
    \subcaption{A simplified relevant usage of methods \texttt{error}, \texttt{empty}, and                  \texttt{flatMap} in \texttt{Mono}}
    \label{fig:rel-usage}
    \end{subfigure}\hfil

    \begin{subfigure}{0.5\textwidth}
\begin{lstlisting}[language=Java]
        return routeId.flatMap(id -> {
            if (routes.containsKey(id)) {
                routes.remove(id);
                return Mono.empty();                               
            }
            return Mono.defer(() -> Mono.error(new NotFoundException("...")));                        
        });                        
\end{lstlisting}
        \subcaption{Method body generated by \tool with DeepSeek Coder, which is exactly matched with the reference}
        \label{fig:rambo-code}
    \end{subfigure}\hfil

    \caption{A MBC problem in repository \texttt{Spring Cloud Gateway}}
    \label{fig:good-example}
\end{figure}

\begin{figure}
 \centering
 \includegraphics[width=1\columnwidth]{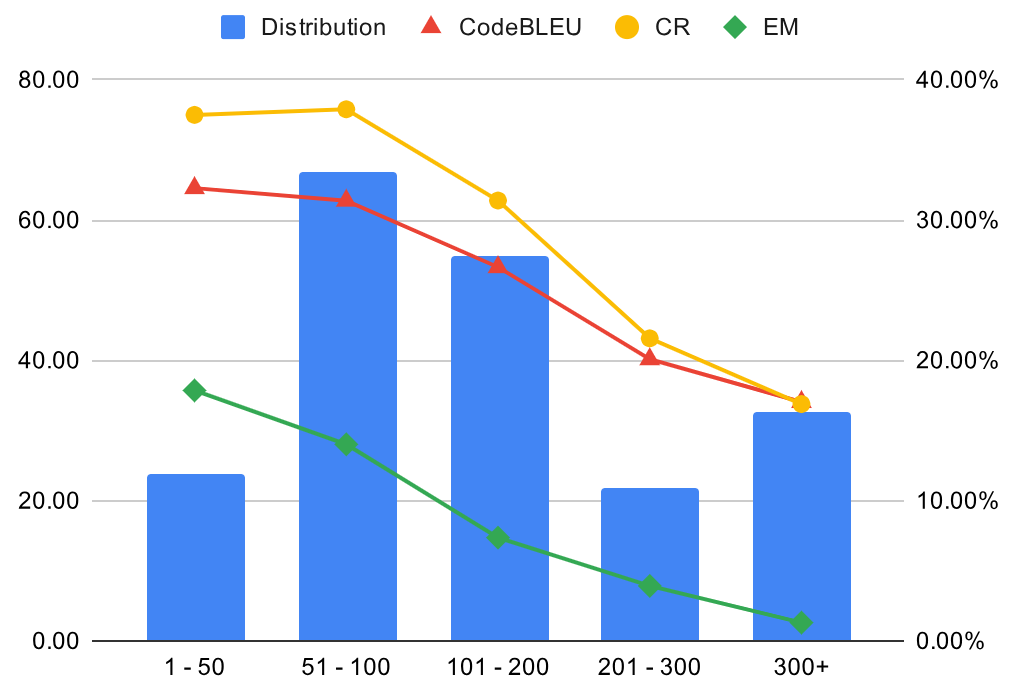}
 \caption{\tool's performance by references' size (LOC), left axis: \textit{CodeBLEU} and \textit{CR}; right axis: \textit{Distribution}}
 \label{fig:RQ1-results-by-body-size}
\end{figure}

Fig.~\ref{fig:RQ1-results-by-body-size} shows \tool's performance across different reference sizes (LOC). As expected, \tool performs much better with smaller code snippets (1--50 LOC), achieving high \textit{CodeBLEU}, \textit{CR}, and \textit{EM}. However, \tool's performance declines gracefully as the reference size increases, particularly for larger codebases (201+ LOC). This trend also appears in the results of the other approaches~\cite{website}.
This decline can be attributed to the increased complexity and dependencies within larger code, which challenge \tool's ability to maintain its accuracy. These findings suggest that while \tool is highly effective for smaller tasks, improvements are needed to enhance its scalability and performance on larger, more complex codebases.

In Tables~\ref{tab:comparison_defect4j} and \ref{tab:comparison_rambo}, the experimental results show that the repo-level MBC approaches perform much better on Defect4J than \tool's benchmark. 
%
This discrepancy is attributed to the fact that \tool's benchmark includes larger repositories and more complex MBC problems with longer reference codes. Additionally, the MBC problems in \tool's benchmark demand an average of 7.2 repo-specific code element invocations (with a maximum of 43), whereas the corresponding problems in Defects4J require only about 4.8 invocations (with a maximum of 17).


\subsection{Answering RQ2: Intrinsic Analysis}
To answer RQ2, we performed the experiments on \tool's benchmark with the state-of-the-art LLM specialized for code, DeepSeek Coder and its variants~\cite{deepseek-coder} as $\mathscr{M}$.

\subsubsection{Impact of Retrieval Technique}

\begin{table}
\centering
\caption{Performance of \tool by Retrieval Techniques}
\label{tab:RQ2-emb}
\scriptsize
\begin{tabular}{lrrrr}\toprule
                        &\textbf{\textit{CodeBLEU}} &\textbf{\textit{CR}}   &\textbf{\textit{EM}} \\\cmidrule{1-4}
Lexical Retrieval                &53.20             &61.70                  &\textbf{18.94} \\
Semantic Retrieval               &52.61             &61.49                  &18.09 \\
Hybrid Retrieval                 &\textbf{53.47}	            &\textbf{62.98}	                &18.51 \\
\bottomrule
\end{tabular}
\end{table}

To investigate the impact of the retrieval technique on \tool's performance, we applied several popular retrieval techniques used to gather relevant context from the repository in \tool: Lexical Retrieval using Byte-Pair Encoding (BPE)~\cite{bpe} with Jaccard similarity, Semantic Retrieval powered by UniXCoder embeddings~\cite{unixcoder} with Cosine similarity, and a Hybrid approach that averages the similarity scores by Lexical Retrieval and Semantic Retrieval.  

Table~\ref{tab:RQ2-emb} compares the performance of \tool across three different retrieval methods. 
As seen, the performance of \tool is slightly impacted by the retrieval techniques with the relative margins of 1.6\% in CodeBLEU, 2.4\% in CR, and 4.8\% in EM. Although Semantic Retrieval offers broader contextual insights, its performance slightly lags behind the other techniques. 
The reason could be that \tool primarily relies on identifying functionally similar methods and code elements by comparing method signatures and element names, where deep semantic understanding may not be as critical.
The Hybrid approach, which combines both lexical and semantic methods, marginally outperforms the individual approaches but introduces additional complexity. Given the small performance gains, this added complexity may not be justified compared to the efficiency of lexical retrieval.
Indeed, Lexical retrieval, which focuses on token-level similarity, stands out for its balance of performance and efficiency. Its lightweight design allows for fast and accurate retrieval of relevant code snippets, making it particularly effective for \tool's MBC tasks.

\subsubsection{EEI and RUE: Contribution Analysis}

To evaluate the impact of \textit{Essential Code Element Identification (EEI)} and \textit{Relevant Usage Extraction (RUE)} on \tool's overall performance, we built two variants of \tool. The first one contains only the component of \textit{Sketch generation} serving as a baseline for evaluating performance without the additional contextual information provided by EEI or RUE. The second variant, \textit{Essential (code) elements only}, analyzes the sketch code to identify most similar elements in the repository and uses their signatures as well as their bodies as the context for code generation by $\mathscr{M}$.

Table~\ref{tab:RQ2-components} shows the performance of these two variants compared to \tool. The \textit{Sketch generation} component serves as a baseline, achieving a \textit{CodeBLEU} of 49.75, a \textit{CR} of 61.06, and an \textit{EM} score of 14.04. 
%
%
%
%
Interestingly, the performance of the \textit{Essential elements only} variant is lower than the sketch-only approach.
Although the essential elements and their bodies are included as context, the performance decrease suggests that introducing them without proper usage context (as in RUE) may mislead the model. The retrieval of functionally similar code elements alone is insufficient without understanding their specific role in the repository, leading to less accurate code generation.

When EEI and RUE are combined in \tool, the LLM achieves the highest performance. The integration of EEI allows \tool to identify essential elements relevant to the method, while RUE provides context for how these elements are used within the repository. This combination ensures that the code LLM receives both relevant code elements and usage patterns, improving its ability to generate functionally correct and syntactically accurate method bodies.

\begin{table}
\centering
\caption{{Contributions of EEI and RUE on \tool's performance}}
\label{tab:RQ2-components}
\scriptsize
\begin{tabular}{crrrr}\toprule
\textbf{Variant} &\textbf{\textit{CodeBLEU}} &\textbf{\textit{CR}} &\textbf{\textit{EM}} \\\cmidrule{1-4}
\textit{Sketch generation}               &49.75 &61.06 &14.04 \\
\textit{Essential elements only} &41.10 &61.49 &13.40 \\
\tool                                   &\textbf{53.20} &\textbf{61.70} &\textbf{18.94} \\
\bottomrule
\end{tabular}
\end{table}

\subsubsection{Impact of Number of Sketches}

To study the impact of sketch generation on \tool's performance, we varied the number of generated sketches in \textit{Sketch Generation} of \tool on Defects4J with DeepSeek Coder as the code LLM.
When using a \textit{single} sketch, \tool achieves a BLEU score of 51.03, an ES of 53.05, and an EM rate of 23.06. Increasing the number of sketches to \textit{five} results in lower scores, e.g., BLEU of 49.43, CodeBLEU of 51.12, and EM of 21.57 in the case of generating 5 sketches. 
In our experiments, we found that that the sketches generated are not substantially different from one another. Additionally, the sketches may even introduce irrelevant elements that slightly degrade performance. Consequently, using a single sketch is sufficient for effectively identifying the essential code elements for method body completion.

\subsubsection{Impact of Relevant Context's Size}

\begin{figure}
 \centering
 \includegraphics[width=1\columnwidth]{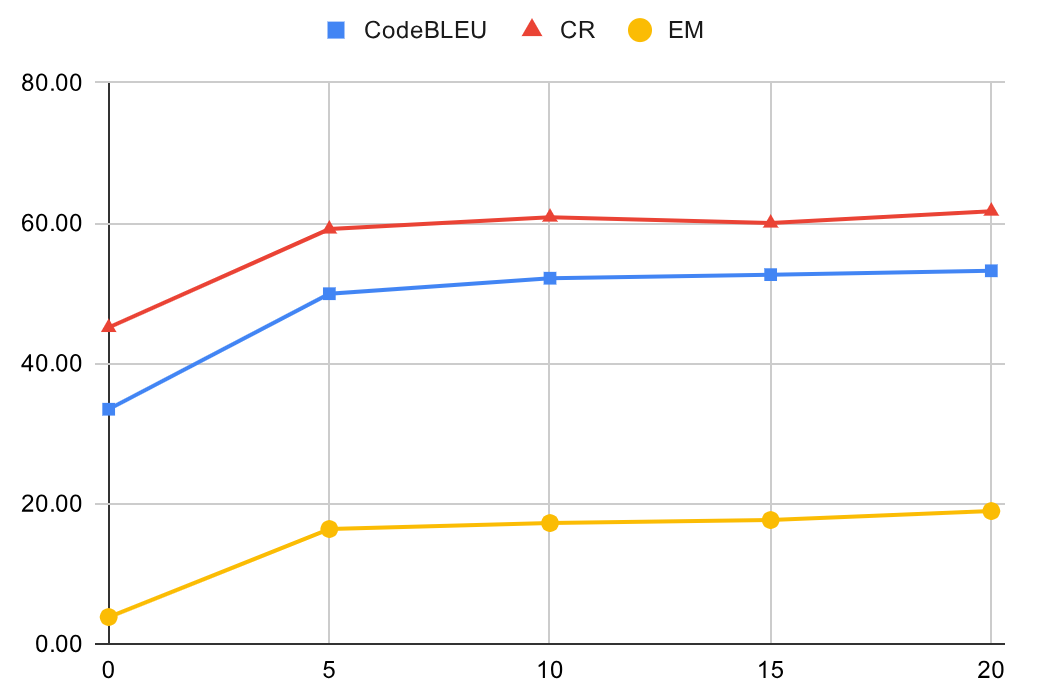}
 \caption{\tool's performance by the size of relevant context, $C_R$ (in the number of methods included in $C_R$)}
 \label{fig:RQ2-CR-size}
\end{figure}

Fig.~\ref{fig:RQ2-CR-size} shows the performance of \tool as the size of the relevant context, $C_R$, increases, measured by the number of relevant methods included.
The performance of \tool improves consistently as more relevant context is incorporated, with notable gains of 59\% in CodeBLEU, 37\% in Compilation Rate, and a 5X increase in Exact Match when the context size increases from 0 to 20 relevant methods. The most substantial improvement occurs when the context size grows from 0 to 5 methods.
However, after a certain threshold, including additional relevant methods in $C_R$ results in diminishing returns. This suggests that \tool can effectively extract essential information with a relatively small set of relevant methods, and beyond that, further context offers marginal benefits.

\subsubsection{Impact of Code LLM's Size}
\begin{figure}
 \centering
 \includegraphics[width=1\columnwidth]{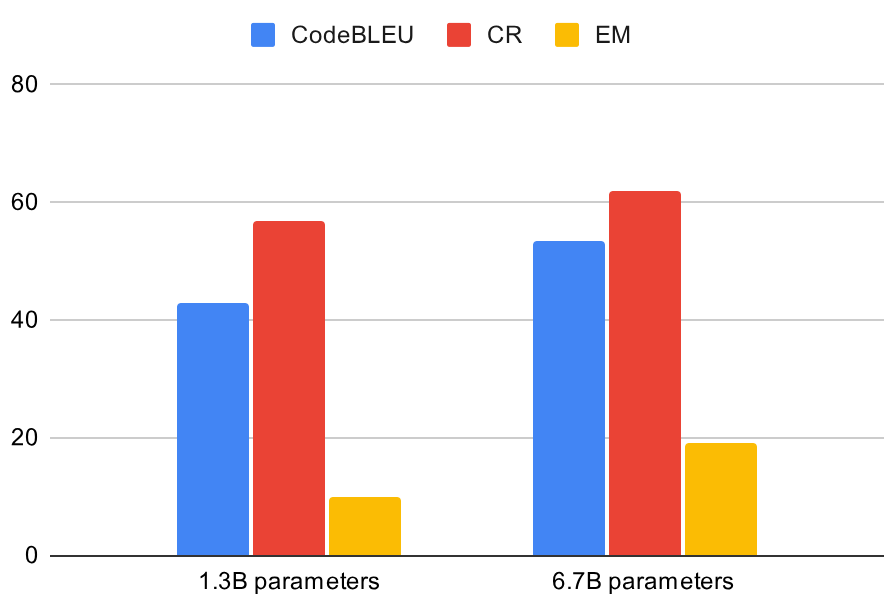}
 \caption{\tool's performance by the sizes of $\mathscr{M}$ (DeepSeek Coder)}
 \label{fig:RQ2-LLM-size}
\end{figure}

To assess how the size of the Code LLM, $\mathscr{M}$, influences \tool's performance, we conducted experiments using two different model sizes. However, our experiments were performed with DeepSeek Coder models of 1.3B and 6.7B parameters due to resource limitations.
As shown in Fig.~\ref{fig:RQ2-LLM-size}, the larger 6.7B model significantly improves \tool's performance, particularly in \textit{CodeBLEU} and \textit{EM}, indicating that a larger model size is advantageous for achieving higher accuracy in repository-level method body completion tasks. 
However, this comes at a cost: code generation time, with the 6.7B model, is 60\% longer compared to \tool with the 1.3B model, highlighting the trade-off between performance and computational efficiency.

\begin{gtheorem} 

\textbf{Answer to RQ2}: \tool's performance improves with relevant context and larger Code LLMs. Lexical retrieval provides the best trade-off between performance and efficiency. Performance gains diminish with excessive relevant context and larger models increase accuracy but slow down generation. 
Generating multiple sketches does not significantly enrich the set of essential code elements, instead, it may introduce redundancy, leading to reduced performance.

\end{gtheorem}


\subsection{Answering RQ3: Sensitivity Analysis}

To answer RQ3, we used \tool with Lexical Retrieval and DeepSeek Coder 6.7B on \tool's benchmark.


\subsubsection{Impact of Surrounding Context Size}
\begin{figure}
 \centering
 \includegraphics[width=1\columnwidth]{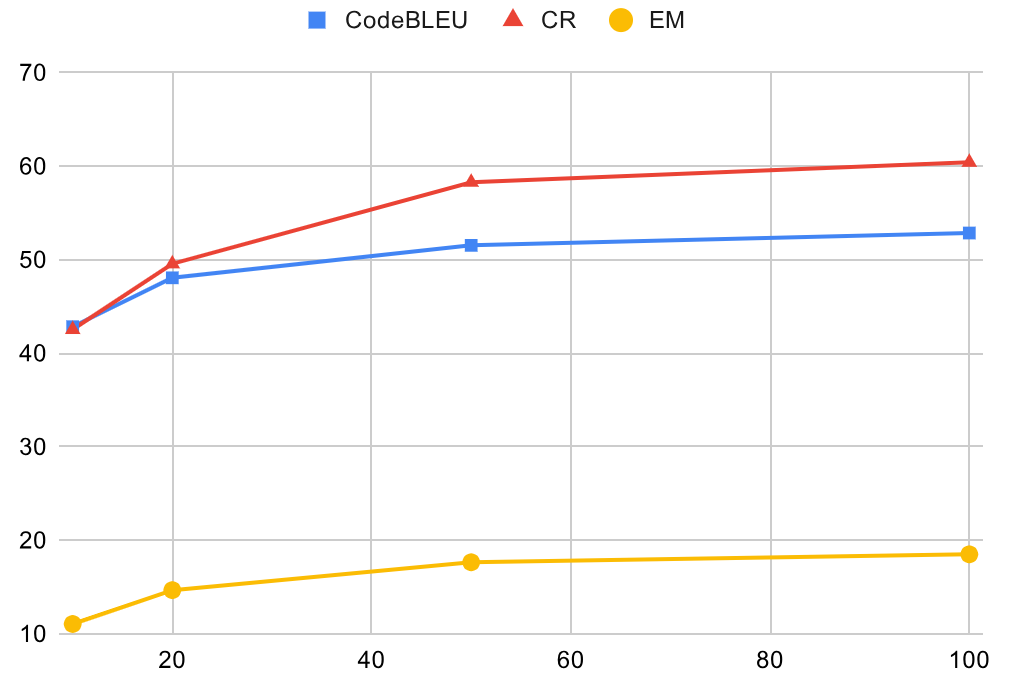}
 \caption{\tool's performance by surrounding context, $C_f$  in different sizes (LOCs)}
 \label{fig:RQ3-Left-context}
\end{figure}

Fig.~\ref{fig:RQ3-Left-context} illustrates how \tool's performance varies with different sizes of the surrounding context, from 10 -- 100 LOCs. 
In this experiment, the surrounding context is structured symmetrically, with the left context and the right context being of identical size.
In general, there is an improvement across all metrics in the performance of \tool when additional surrounding context is provided. Particularly, when the surrounding context size increases from 10 -- 20 LOCs on both sides, \tool's performance significantly improves, 12\% in CodeBLEU, 16\% in CR, and 33\% in EM. As the context size increases further (20--50 LOCs), the improvements remain steady but moderate. From 50--100 LOCs, the improvement trend slows down. Beyond 100 LOCs, a plateau effect becomes apparent. This shows that additional surrounding context might introduce redundancy or irrelevant information, which does not contribute meaningfully to generating high-quality method bodies. 

Overall, these results demonstrate the importance of balancing surrounding context size in \tool. For practical applications, one should limit the surrounding context to around 50 LOCs, as larger contexts yield only marginal improvements in \tool's performance while adding complexity and potential noise to the generation process.

\subsubsection{Impact of Repository Size}
\begin{figure}
 \centering
 \includegraphics[width=1\columnwidth]{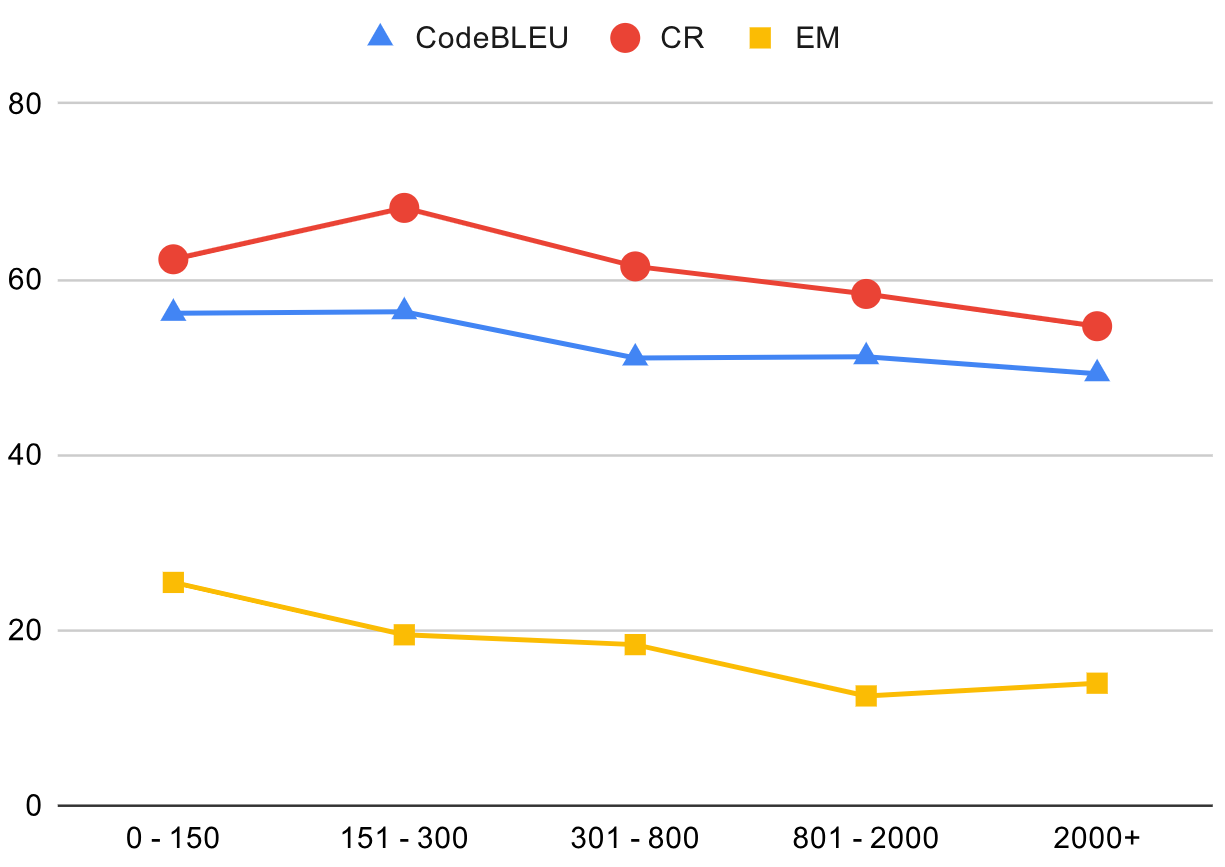}
 \caption{\tool's performance by repository sizes (files)}
 \label{fig:RQ3-repo-size}
\end{figure}

Fig.~\ref{fig:RQ3-repo-size} illustrates \tool's performance across repositories of various sizes. \tool performs optimally on smaller repositories (20 out of 26 repositories with fewer than 800 files). On average, it achieves a CodeBLEU score of 54.5, a Compilation Rate (CR) of 64\%, and an Exact Match (EM) rate of 21.1\%.
However, as the repository size increases, there is a decline in performance across all metrics. In particular, for very large repositories containing more than 2,000 files, CodeBLEU drops to 49.23, and EM decreases to 13.95\%.
These findings suggest that while \tool excels in smaller repositories, its ability to maintain high accuracy with larger and more complex repositories.


\subsubsection{Impact of Programming Languages}

\begin{table}
\centering
\caption{{Repo-level MBC performance} \tool {and the others on RepoEval dataset (Python)}}
\label{tab:RQ3-python}
\scriptsize
\begin{tabular}{lrrrrr}\toprule
&BLEU &CodeBLEU &Pass@1 &EM \\\cmidrule{1-5}
\cellcolor[HTML]{ffe599}\tool Oracle &\cellcolor[HTML]{ffe599}56.09 &\cellcolor[HTML]{ffe599}52.58 &\cellcolor[HTML]{ffe599}46.33 &\cellcolor[HTML]{ffe599}20.11 \\
\cellcolor[HTML]{ffe599}RepoCoder Oracle &\cellcolor[HTML]{ffe599}52.71 &\cellcolor[HTML]{ffe599}48.91 &\cellcolor[HTML]{ffe599}40.47 &\cellcolor[HTML]{ffe599}15.28 \\
Infile &35.82 &34.22 &24.63 &5.90 \\
RawRAG &48.21 &44.73 &36.66 &14.21 \\
RLCoder &52.79 &50.38 &43.70 &18.23 \\
RepoCoder &51.08 &48.10 &40.18 &16.89 \\
\tool &\textbf{55.78} &\textbf{52.39} &\textbf{46.92} &\textbf{20.11} \\
\bottomrule
\end{tabular}
\end{table}

Table~\ref{tab:RQ3-python} shows the performance of \tool with Code Llama on RepoEval which is a Python benchmark proposed simultaneously with RepoCoder~\cite{repocoder}.
The results on this Python benchmark indicate that \tool maintains significant improvements of 19\%--240\% in EM and up to 90\% in Pass@1 over the existing approaches, even their upper-bound performance. These demonstrate \tool's ability to generate accurate and contextually relevant method bodies regardless of language. These findings affirm that the strategies employed by \tool, particularly the identification of essential code elements and relevant usage extraction, are robust and effectively generalize across diverse programming environments.

\begin{gtheorem} 
\textbf{Answer to RQ3}: \tool performs best with balanced surrounding context ($C_f$) and excels in smaller repositories. Performance declines in larger repositories due to increased complexity and generation times. 
Additionally, our experiments on both Java and Python demonstrate that \tool's approach generalizes effectively across programming languages.
\end{gtheorem}


\subsection{Answering RQ4: Time complexity} 

For the retrieval tasks, \tool required an average of 8 minutes to extract all elements from a repository. The process of converting code sequences into bag-of-words embeddings took approximately 4.4 seconds each repository. Meanwhile, \tool took around 0.9 seconds per sample to search the top-\textit{k} functionality-similar and relevant usages during EEI and RUE.
Additionally, the generation tasks in \tool's pipeline, including sketch and method body generation, took about 17.7 seconds on average.


\subsection{Threats to Validility}
The main threats to the validity of our work consist of internal, construct, and external threats.

\textbf{Threats to internal validity} include the hyperparameter settings of the baseline methods and \tool. To mitigate this threat, we systematically varied the settings of \tool and thoroughly examined its performance (as discussed in Section~\ref{sec:results}). 
Additionally, we reused the implementations and settings from the original papers of the baseline approaches~\cite{repocoder} to ensure consistency in comparison. Another potential threat arises from the methods used to analyze generated sketch code and identify essential elements within the repository. To address this, we employed a robust code analyzer~\cite{joern} and popular retrieval mechanisms applied by existing studies~\cite{repocoder,repoformer,repohyper}, ensuring our approach aligned with established practices. 
Furthermore, to minimize the threat of implementation errors, we conducted a detailed review of our code and have made it publicly available~\cite{website} to facilitate replication and validation by other researchers.

\textbf{Threats to construct validity} relate to the appropriateness of our evaluation procedure. We utilized widely accepted metrics for both text-based and code-based evaluations, including \textit{BLEU}, \textit{CodeBLEU}, \textit{Exact Match}, \textit{Compilation Rate}, and \textit{pass@k}, which are standard measures in code completion and code generation research. 
A potential threat remains due to the absence of a human evaluation, which could provide a more nuanced assessment of our approach. To address this limitation, we plan to incorporate human evaluations in future work, wherein experienced developers will assess the generated method bodies based on criteria such as functionality, accuracy, and adherence to repository standards. 
Additionally, the controlled environments used in our experiments may not fully represent real-world scenarios. To mitigate this, we conducted evaluations in diverse settings and examined performance across multiple scenarios and configurations.

\textbf{Threats to external validity} primarily concern the generalizability of our results to different repositories and programming languages. To reduce this threat, we selected representative code LLMs models and repositories widely recognized in both NLP and SE communities~\cite{deepseek-coder,gpt,codellama}. 
Another potential threat is using a limited set of Java and Python repositories in our experiments. Our current study focuses on a specific set of repositories, which may limit the applicability of our findings to other types of codebases or programming languages. In future work, we intend to extend our experiments to include a broader range of repositories and languages to enhance the generalizability of our results.


\section{Related Work}

\tool is closely related to the research on \textbf{repository-level code completion}, which utilizes the broader context provided by an entire code repository~\cite{MGD, cocomic, codeplan, RepoPrompts, repocoder,repoformer,rlcoder, repohyper}. 
RepoPrompts~\cite{RepoPrompts} introduces an approach that first generates prompts reflecting the repository's structure and predefined relevant contexts, then learns to select the prompt most likely to produce the desired output.
In the work by Wang~\etal~\cite{rlcoder}, the task of retrieving useful context is framed as a reinforcement learning problem, where the retriever parameters are iteratively updated.
MGD~\cite{MGD} monitors the code generated by LLMs, queries static analysis tools in the background, and leverages the resulting information to guide the LLMs during the decoding stage.
RepoCoder~\cite{repocoder} integrates a similarity-based retriever and a code LLM into an iterative retrieval-generation pipeline.
RepoHyper~\cite{repohyper} improves the context retrieval quality over the prior work by introducing a hypergraph-based retriever that learns to model cross-file dependencies. While RepoHyper demonstrates strong performance in selecting informative context for code generation, it operates under the assumption that the necessary imported files are already present in the file being completed. Furthermore, RepoHyper requires a  training phase for re-ranking and selecting suitable retrieved contexts, which may limit its applicability in dynamic or unseen repositories.

Although these efforts show promising performance, \tool differs from them in that \tool requires neither training nor modifications to the underlying code LLMs. 
To further improve its practical usability, \tool makes no assumptions about the existence of imported files in the current file.
Additionally, instead of similarity-based retrieval, \tool emphasizes the extraction of code elements most relevant to the infilling method and their usage contexts. This design ensures that the generated method bodies are not only contextually accurate but also aligned with the repository’s specific coding practices.

\textbf{Code completion} is a critical feature in modern Integrated Development Environments (IDEs) and has garnered significant interest from researchers~\cite{rule-based-cc1, zhou2022improving, codefill, hellendoorn2017deep, bigcode, ase19, arist,intellicode,reacc,docprompting,syntax-guided,llm-lib,ReCode}. 
Earlier approaches~\cite{rule-based-cc1,rule-based-cc2,cpathminer} relied on rule-based techniques or code example patterns for completing code.
In recent years, learning-based approaches have been developed to enhance code completion performance~\cite{naturalness, tu2014localness, zhou2022improving, codefill, hellendoorn2017deep, bigcode, ase19, arist,intellicode}.
The advent of large language models (LLMs) has further advanced this field, leading to their integration into code completion and generation tasks~\cite{reacc, docprompting, syntax-guided, llm-lib, ReCode}.
For instance, ReACC~\cite{reacc}, a retrieval-augmented code completion framework inspired by coding practices, enhances code generation by incorporating lexical copying and semantic retrieval, followed by generation through a decoder-only transformer model.
ReCode~\cite{ReCode} advances neural code generation by integrating subtree retrieval from existing code examples.
DocPrompting~\cite{docprompting} improves code generation by leveraging code documentation, effectively addressing challenges in generating code for unfamiliar functions and libraries.
Zhang~\etal~\cite{syntax-guided} propose kNN-TRANX that improves code generation by leveraging syntax constraints for retrieval, reducing noise and computational time.

Recently, several \textbf{learning-based approaches} have been proposed for specific Software Engineering tasks such as code translation/migration~\cite{pan2024lost,zhang2023multilingual,structcoder,lost-in-transl, icse2022, icpc2019}, text-code translation~\cite{patchexplainer,generate_log_2024,comment_gen_2024, gao2023code, icse20}, test generation~\cite{liu2023fill, lan2024deeply}, bug detection~\cite{oppsla19, CodeJIT, qiu2024vulnerability,nguyen2024context}, bug fix identification~\cite{kse}, and program repair~\cite{xia2023automated,ruan2024timing}.


\section{Conclusion}

In this paper, we introduced \tool, a novel Retrieval-Augmented Generation (RAG)-based approach designed specifically for repository-level method body completion (MBC). Our approach centers on the accurate identification and utilization of essential repo-specific code elements, significantly enhancing the performance of code LLMs in generating contextually relevant and syntactically correct method bodies.
Our experimental results show \tool's superiority across all key metrics, including \textit{BLEU}, \textit{CodeBLEU}, \textit{Compilation Rate}, pass@$k$, and \textit{Exact Match}. Notably, \tool not only surpassed existing methods but also exceeded the performance of RepoCoder Oracle, setting a new benchmark in the field.
These show \tool's robustness, accuracy, and practical utility in real-world code generation tasks. By leveraging essential code elements and relevant usages, \tool addresses the limitations of existing similarity-based approaches, providing a more precise and reliable solution for MBC in complex repositories.


\section*{Acknowledgement}
This research is supported by Vietnam National Foundation for Science and Technology Development (NAFOSTED) under grant number 102.03-2023.14.
This research is also partly supported by OpenAI's Researcher Access Program.

\bibliographystyle{elsarticle-num}

\bibliography{references}

\end{document}